\documentclass[prd,floatfix,showpacs]{revtex4}
\date{\today}
\usepackage{amsthm}
\usepackage{amsmath}
\usepackage{amsfonts}
\usepackage{graphicx,epsfig}
\newcommand{\bp}{\mathbf{p}}

\newcommand{\bk}{\mathbf{k}}

\renewcommand{\lnot}{\!\!\not{\;l}}

\newcommand{\nn}{\nonumber}

\newcommand{\ba}{\begin{eqnarray}}
\newcommand{\ea}{\end{eqnarray}}
\newcommand{\ie}{{\it i.e.,\ }}
\newcounter{bean}
\newcounter{pea}

\newtheorem{theorem}{Theorem}

\newtheorem{statement}{Statement}

\reversemarginpar

\newcommand{\E}{{E}}
\newcommand{\N}{I}
\newcommand{\gammaE}{\gamma_{E}}

\begin{document}
\title{An operator representation for Matsubara sums}

\author{Olivier Espinosa}
\email{olivier.espinosa@fis.utfsm.cl}
\author{Edgardo Stockmeyer\footnote{Present Address: Mathematisches Institut
  LMU M\"{u}nchen , Theresienstr. 39,
  80333 M\"{u}nchen, Germany.}
}
\email{stock@mathematik.uni-muenchen.de}
\affiliation{Departamento de F{\'\i}sica, Universidad T{{\'e}}cnica Federico
Santa Mar{\'\i}a, Casilla 110-V, Valpara{\'\i}so, Chile}

\begin{abstract}

In the context of the imaginary-time formalism for a scalar
thermal field theory, it is shown that the result of performing
the sums over Matsubara frequencies associated with loop Feynman
diagrams can be written, for some classes of diagrams, in terms of
the action of a simple linear operator on the corresponding energy
integrals of the Euclidean theory at $T=0$. In its simplest form
the referred operator depends only on the number of internal
propagators of the graph.

More precisely, it is shown explicitly that this \emph{thermal
operator representation} holds for two generic classes of
diagrams, namely, the two-vertex diagram with an arbitrary number
of internal propagators, and the one-loop diagram with an
arbitrary number of vertices.

The validity of the thermal operator representation for diagrams
of more complicated topologies remains an open problem. Its
correctness is shown to be equivalent to the correctness of some
diagrammatic rules proposed a few years ago.
\end{abstract}

\maketitle

\section{Introduction}
\label{introduction}

In the imaginary-time formalism, the calculation of a loop diagram
in quantum field theory at finite temperature necessarily involves
sums over Matsubara frequencies\cite{books-on-FTQFT}, an operation
that we shall generically call \emph{the Matsubara sum} associated
with the graph. Although this sum can be computed in a number of
ways, usually in a systematic fashion, such computations can
become quite tedious for higher loop
diagrams\cite{saclay-method,Guerin}.

In reference \cite{DES-97} a set of simple diagrammatic rules were
postulated to write down an explicit expression for the
\emph{result} of performing the Matsubara sum associated with any
finite-temperature Euclidean Feynman graph (in a scalar theory).
Because of the similitude of the diagrammatic expansion with the
one associated with the non-covariant old-fashioned perturbation
theory formalism (at zero temperature), these diagrammatic rules
will be referred to as the OFPT-rules.

Although in reference \cite{DES-97} the OFPT-rules were explicitly
\emph{verified} to hold for a few nontrivial diagrams, they were
presented as a sort of empirical discovery, with no rigorous proof
given.
\medskip

In this paper we restate the diagrammatic analysis of reference
\cite{DES-97} in an algebraic rather than diagrammatic fashion,
and extend its validity to two particular \emph{classes} of
diagrams, to be described below. For these diagrams we establish
that the full result of performing the Matsubara sum associated
with a given Feynman graph can be completely determined from its
zero-temperature counterpart, by means of a simple linear
operator, as shown in \eqref{eq:main-conjecture} below. We have
termed this result the \emph{thermal operator representation}
(TOR) of the Matsubara sum.

The two classes of diagrams for which we have been able to prove
the correctness of the TOR are: (a) diagrams with
two vertices and an arbitrary number of scalar internal
propagators; and (b) one-loop diagrams with $I$ vertices and $I$
scalar internal propagators, with $I\ge 1$.
In what follows, whenever we refer to a Feynman
diagram we implicitly assume that the diagram actually belongs to
one of the classes just described, except when specifically
qualified otherwise.

The precise mathematical formulation of the thermal operator
representation is presented in the next section. Leaving out many
of the technicalities, its contents is as follows. Consider the
Matsubara sum of a (amputated) scalar loop Feynman graph with $\N$
internal lines and external Euclidean 4-momenta
$P_\alpha=(p_{\alpha}, {\bf p}_\alpha)$. [A word about the
notation: in order to avoid clutter, we will omit the customary
$0$ superscript on Euclidean energy variables. Since we shall not
denote in this paper the modulus of a 3-momentum vector $\bf p$
with the corresponding italic symbol $p$, there should be no
danger of confusion]. Instead of following the usual practice of
parameterizing all internal line 4-momenta in terms of a few
independent loop 4-momenta, by explicitly requiring 4-momentum
conservation at each vertex, we choose to assign each internal
line independent 3-momentum ${\bf k}_i$ and Matsubara frequency
$k_i$ and impose 4-momentum conservation by means of an
appropriate number of delta functions. In this form, the Matsubara
sum will depend only on the external Euclidean energies
$p_{\alpha}$ (which enter through Kronecker delta functions
enforcing energy conservation at each vertex), on the kinematic
energies of the internal lines, $E_i:=({\bf k}_i^2+m_i^2)^{1/2}$
appearing in the propagators and, of course, the temperature $T$.
Since there is no explicit dependence of the Matsubara sum on
spatial 3-momenta, external or internal, we shall suppress all
reference to these in this paper, whenever possible.

Let the unsubscripted symbols $p$ and $E$ denote, respectively,
the full set of Euclidean external and kinematic internal
energies, $p:=\left\{p_1, p_2, \dots, p_{n}\right\}$ and
$E:=\left\{E_1, E_2, \dots, E_I\right\}$. Now, if we introduce the
\emph{Matsubara $D$-function} of the graph, $D(p, E, T)$,
essentially as the Matsubara sum multiplied by the product of all
internal kinematic energies, then we claim that
\begin{equation}\label{eq:main-conjecture}
D(p, E, T)=\hat{O}(E, T)D_0(p, E),
\end{equation}
where $\hat{O}(E, T)$ (whose explicit form we give in the next
section) is a linear operator that depends on the topology of the
diagram but is independent of the external Euclidean energies
$p$. The object acted upon by this operator, $D_0(p, E)$, is
simply the corresponding $D$-function for the Euclidean
zero-temperature graph, $D_0(\omega, E)$, defined for real and
continuous external energies $\omega:=\left\{\omega_1,
\omega_2,\ldots,\omega_n\right\}$, evaluated at $\omega=p$:
\begin{equation}\label{eq:D0-evaluated-at-p}
D_0(p,E)=D_0(\omega, E)
\Big|_{\omega=p}.
\end{equation}
We shall call
$\hat{O}(E,T)$ the (Euclidean) \emph{thermal operator}.

As we shall see in the next section, the thermal operator has a
form that can be readily and naturally extrapolated to diagrams of
arbitrary topologies. Although this makes it tempting to
conjecture that the thermal operator representation holds for
completely arbitrary diagrams, this remains an open problem and
more work is needed to settle the issue.

However, if true in general, the representation
\eqref{eq:main-conjecture} would have several immediate important
consequences: (a) it would show that the full finite temperature
result is encoded in the zero-temperature function $D_0(p, E)$,
rendering the actual computations of the Matsubara sums completely
unnecessary;
(b) since all dependence on external energies $p$ is contained
into the zero-temperature function $D_0$, any analytic
continuation of $D(p,E,T)$ to complex values of the external
energies, physically meaningful or not, would need only be carried
out on $D_0$. By the same token, the study of imaginary parts of
analytically continued Euclidean Green functions, \ie the subject
of cutting rules, would need only be done at the level of the
zero-temperature function $D_0$, since the thermal operator is
real (we give an example of this in the last section of this
paper).
(c) since the thermal operator is bounded as the internal energies
$E_i$ tend to infinity, it would be enough to renormalize $D_0$ in
order to renormalize the full finite temperature result. This is
consistent with a well-known result in renormalization of thermal
field theories.

Although there have appeared in the literature several works that
touch upon the relationship between the full calculation of
finite-temperature Feynman graphs and their zero-temperature
counterparts (usually interpreted in terms of forward scattering
amplitudes in vacuum), both in the Euclidean imaginary-time
\cite{Frenkel et.al.} and in the real-time formalisms\cite{Brandt
et.al.}, we are unaware of any discussion of a representation of
the simple form \eqref{eq:main-conjecture}, as given here.

We emphasize that all the results presented in this paper are
formulated in the context of the Euclidean imaginary-time
formalism, and we will have nothing to say here about their
relationship to or consequences for the real-time formalism,
except for the remark made above about the possible analytic
continuations of the Euclidean Green functions to complex values
of the Euclidean external energies. The latter subject has been
studied at great length in the literature\cite{analytic
continuations}, along with the connection between different
analytically continued Euclidean Green functions and the retarded,
advanced or time-ordered Green functions of the real-time
formalism, and the subject of cutting rules in the real-time
formalism\cite{cutting rules}.
\medskip

The structure of the paper is as follows: In section
\ref{sec:conjecture} we shall present the general form of the
thermal operator representation (TOR) for the Matsubara sum of a
general scalar graph, in two alternative forms. In sections
\ref{sec:tadpole-diagram}, \ref{sec:two-vertex-diagram} and
\ref{sec:one-loop-diagram} we prove that the TOR holds,
respectively, for a one-loop single-propagator tadpole-like graph,
for a generic graph with two vertices, and for a generic one-loop
graph; the number of internal propagators is allowed to be
arbitrary (but at least equal to two) in the last two cases.
Additional supporting evidence for the validity of the TOR for
graphs of arbitrary topologies and our conclusions are given in
section \ref{sec:conclusions}. The reformulation of the
old-fashioned perturbation theory rules of reference \cite{DES-97}
in the form of the present representation has been relegated to an
appendix.

\medskip

\section{A representation for the Matsubara sum}\label{sec:conjecture}

In a scalar field theory, the mathematical expression
corresponding to an amputated graph with $n+1$ vertices
($n\geq 1$), $\N$ internal lines, and external 4-momenta
$P_\alpha=(p_{\alpha}, {\bf p}_\alpha)$ has the form
\begin{equation}
\frac{(-\lambda)^{n+1}}S\int
[\prod\limits_{i=1}^{\N}\frac{d^3k_i}{\left( 2\pi \right)^3 2E_i}%
\prod\limits_{V=1}^n (2\pi )^3\delta^{(3)} ({\bf k}_V)]
\;D(p, E, T),
\label{eq:D-def}
\end{equation}
where $\lambda$ represents the coupling constant and $S$ is the
symmetry factor of the graph; ${\bf k}_i$ is the spatial
3-momentum of the $i$-th internal line and $E_i:=({\bf
k}_i^2+m_i^2)^{1/2}$ is its associated kinematic energy; ${\bf
k}_V$ denotes the total 3-momentum entering vertex $V$; the
unsubscripted symbols $p$ and $E$ denote, respectively, the full
set of Euclidean external and kinematic internal energies,
$p:=\left\{p_1, p_2, \dots, p_{n}\right\}$ and $E:=\left\{E_1,
E_2, \dots, E_I\right\}$; and $T$ is the temperature.
The delta functions ensure conservation of spatial 3-momentum at
each vertex, so that the integration measure reduces essentially
to an integration over the 3-momenta of the $L=\N-n$ independent
loops. In the finite temperature Euclidean formalism all lines,
external and internal, carry discrete Euclidean energies which are
integer multiples of $2\pi T$. Each internal line has an
associated Matsubara frequency, denoted by $k_j = 2\pi T n_j$.
The $D$-function is given by the normalized Matsubara sum
\begin{equation}\label{def-D-function}
D(p, E, T)= \gammaE \, T^{L}\sum_{n_1, n_2, \dots, n_\N}
\prod_{j=1}^\N \Delta(k_j, E_j)\delta(p, k),
\end{equation}
where
\ba\label{eq:def-gamma}
\gammaE:=\prod_{i=1}^{\N}2E_i,
\ea
$L$ is the number of independent loops in the graph, and
$\Delta(k_j,E_j)$ is the scalar propagator associated with the
$j$-th internal line, with
\ba\label{eq:def-Delta}
\Delta(k,E):=\frac{1}{k^2+E^2}.
\ea
The sums over each $n_j$ run from $-\infty$ to $+\infty$. The
$\delta$-function, with $k=\left\{k_1, \dots, k_I\right\}$, is a
generalized Kronecker delta which ensures conservation of energy
at each vertex. The topology of the diagram is totally contained
in this generalized delta.

The OFPT-rules given in \cite{DES-97}, which are reproduced in
appendix \ref{ofpt-rules-statement}, were conjectured to allow us
to write down the complete result for \eqref{def-D-function} by a
simple diagrammatic analysis. But as shown in appendix
\ref{operator-approach}, there exists a simple algebraic
representation for the diagrammatic OFPT-rules, so that the
conjecture of reference \cite{DES-97} can be recast in the
following terms:
\bigskip

\begin{statement}\label{conjecture-1}
\noindent\textbf{[Thermal Operator Representation]}
The $D$-function defined in \eqref{def-D-function} for an
amputated Feynman graph can be expressed in the form
\begin{equation}\label{eq:conjecture-1}
D(p, E, T)=\hat{O}(E, T)D_0(\omega, E)
\Big|_{\omega=p},
\end{equation}
where $D_0(\omega, E)$ is the $D$-function of the
Euclidean zero-temperature graph and $\hat{O}(E, T)$, the
thermal operator, is the following linear operator:
\begin{equation}\label{eq:thermal-operator-form1}
\begin{split}
\hat{O}(E,T):=1+&\sum_{i=1}^I  n_i(1+\mathcal{S}_i)
+\sideset{}{^\prime}\sum_{\langle i_1,  i_2\rangle} n_{i_1}n_{i_2}
(1+\mathcal{S}_{i_1})(1+\mathcal{S}_{i_2})\\
+& \dots + \sideset{}{^\prime}\sum_{\langle i_1, \dots,  i_{L}\rangle}
\prod_{l=1}^{L} n_{i_l}(1+\mathcal{S}_{i_l}).
\end{split}
\end{equation}
Here $n_i\equiv n(E_i)$, where
$n(E)=\left(e^{\beta E}-1\right)^{-1}$ is the Bose-Einstein
thermal occupation factor; $S_i:=S_{E_i}$ is a reflection
operator, $S_x f(x):=f(-x)$; the indices $i_1, i_2, \ldots$ run
from 1 to $\N$ (the number of internal propagators) and the symbol
$\langle i_1, \dots,  i_k\rangle$ stands for an unordered
$k$-tuple with no repeated indices, representing a particular set
of internal lines. The primes on the summation symbols imply that
certain tuples $\langle i_1, \dots, i_k\rangle$ are to be excluded
from the sums: those such that if we snip all the lines
$i_1,\dots,i_k$ then the graph becomes disconnected.
\end{statement}

Note that the operator $\hat{O}(E,T)$ contains products of at most
$L$ thermal occupation factors $n(E_i)$, since for a $L$-loop graph
the maximum number of lines that can be snipped without
disconnecting the graph is precisely $L$. This generic feature of
the thermal graph in the imaginary-time formalism is of course
well known. However, as discussed in sections
\ref{sec:two-vertex-diagram} and \ref{sec:one-loop-diagram}, there
exists a simpler algebraic form for the thermal operator: \medskip

\begin{statement}\label{conjecture-2}
\noindent\textbf{[Simpler form of the Thermal Operator]}
When acting on the zero-temperature $D$-function, $D_0(p,
E)$, the thermal operator $\hat{O}(E, T)$ can be
replaced by the the simpler
\begin{equation}\label{eq:thermal-operator-form2}
\hat{O}_\star(E, T)=\prod_{i=1}^I [1+n_i(1+\mathcal{S}_i)].
\end{equation}
\end{statement}

Note that the operator $\hat{O}_\star(E, T)$ in
\eqref{eq:thermal-operator-form2} can be expanded as in
\eqref{eq:thermal-operator-form1}, except that the summation
symbols carry no primes, that is, all tuples $\langle i_1, \dots,
i_k\rangle$ ($1\le k \le \N$) are allowed in the sum.
Clearly, the form  \eqref{eq:thermal-operator-form2}
will follow from \eqref{eq:thermal-operator-form1} if we can
somehow show that tuples associated with disconnected graphs (the
ones excluded from the summations in
\eqref{eq:thermal-operator-form1}) give rise to operators that
produce a vanishing contribution to the $D$-function in
\eqref{eq:conjecture-1}. So, the simpler representation will follow
from \eqref{eq:thermal-operator-form1} if the following statement
is true:
\bigskip

\begin{statement}\label{conjecture-3}
\noindent\textbf{[Cut sets do not contribute]}
The zero-temperature $D$-function,
$D_0(\omega, E)$, is annihilated by the operators
\begin{equation}\label{annihilation-operator}
\mathcal{A}(C):=\prod_{i_l\in C}(1+\mathcal{S}_{i_l}),
\end{equation}
where $C$ stands for a \emph{cut set} of the graph, that is,
any set of indices $i_1,\ldots,i_k$ such that the graph becomes
disconnected if the corresponding lines are snipped.
\end{statement}
\medskip

We make clear at this point that, although we make
reference to \emph{cut sets}, we imply no connection to the
concepts of cuts and cut diagrams as they are usually understood
in diagrammatic quantum field theory. Cut sets are determined solely
by the topology of the diagram, and have no further mathematical
or physical meaning.
\medskip

The goal of the next three sections is to prove that these
statements are indeed true for the two generic types of graphs
described in the introduction. The strategy  of the proof will be
to evaluate the Matsubara sums contained in $D(p, E, T)$ by
conventional means, namely the contour integration method or the
Saclay method, and then show that the result can be written as in
the right-hand side of \eqref{eq:conjecture-1}.

\medskip

\section{The simplest loop diagram}
\label{sec:tadpole-diagram}

\begin{figure}
\centerline{\epsfig{file=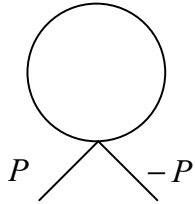,height=4cm,angle=0}}
\caption[]{A one-loop single-propagator diagram.}
\label{fig:tadpole}
\end{figure}
We begin by considering a one-loop graph with only one internal
propagator, as the one shown in figure \ref{fig:tadpole}. This
particular graph contributes at first order to the self-energy in
the $\lambda\phi^4$ theory. The actual number of external legs of
the graph is unimportant, since we are only interested in the
Matsubara sum associated with the loop. Although we could have
considered this graph as the simplest case of the generic one-loop
graph considered in section \ref{sec:one-loop-diagram}, we prefer
to analyze it separately, since the proof given in section
\ref{sec:one-loop-diagram} applies more naturally to the case of
two or more internal propagators.

According to \eqref{def-D-function} the $D$-function for the graph
of figure \ref{fig:tadpole} is simply given by
\ba
D(p,E,T) = (2E)\,T\sum\limits_{n=-\infty}^{+\infty}
\frac{1}{(2\pi T n)^2+E^2}.
\ea
The sum above can be computed in a variety of ways and the result
is well known\cite{books-on-FTQFT}. One obtains
\ba\label{eq:D-function-result-tadpole}
D(p,E,T) = 1 + 2n(E),
\ea
where $n(E)=\left(e^{\beta E}-1\right)^{-1}$ is the Bose-Einstein
thermal occupation factor.
The zero-temperature $D$-function, $D_0(\omega,E)$ can be
computed directly from its definition,
\ba\label{eq:def-D0-tadpole}
D_0 (\omega,E) = (2E)\int\limits_{ - \infty }^\infty  {\frac{{dk_0 }}
{{2\pi }}} \frac{1}
{{k_0^2  + E^2 }},
\ea
or simply by taking the limit $T\to 0$ of $D(p,E,T)$ in
\eqref{eq:D-function-result-tadpole}. The result is
\ba
D_0 (\omega,E) = 1.
\ea
Since a constant function is unchanged by the reflection
operator $\mathcal{S}_{E}$ defined by
\ba\label{eq:def-reflection-operator}
\mathcal{S}_{E}f(E):=f(-E),
\ea
where $f$ is any regular function in the variable $E$, we
certainly have the identity
\ba
D(p,E,T) = \big[ 1 + n(E)(1+\mathcal{S}_{E}) \big]D_0
(p,E),
\ea
which proves that the thermal operator representation
given by \eqref{eq:conjecture-1} and \eqref{eq:thermal-operator-form1}
does hold for the simple graph we are considering.

\section{The two-vertex diagram}
\label{sec:two-vertex-diagram}
\subsection{Calculation}\label{saclay-method}
\begin{figure}
\centerline{\epsfig{file=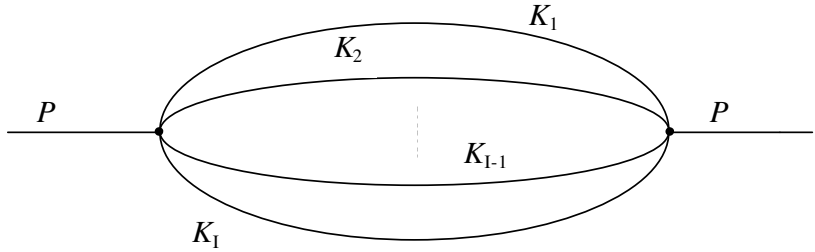,height=4cm,angle=0}}
\caption[]{A two-vertex diagram with $I$ internal lines.}
\label{fig:2-vertex}
\end{figure}

The Matsubara sum for the two-vertex diagram with $\N$ internal
propagators shown in figure \ref{fig:2-vertex} is most
conveniently calculated using the Saclay
method\cite{saclay-method}, which we now briefly review.

Let $K:=(k, \mathbf{k})$ be the Euclidean 4-momentum vector
associated with a given internal line; $k$ is a Matsubara
frequency to be summed over.

Then each scalar propagator,
\ba\label{def:scalar-propagator}
\Delta(K):=\frac{1}{K^2+m^2}=\frac{1}{k^2+E^2}:=\Delta(k,E),
\ea
where $E:=E_{\bk}=\sqrt{\bk^2+m^2}$, is represented as
\begin{equation}\label{propagator-saclay-rep}
\Delta(k,E)=\int_0^\beta d\tau \,e^{ik\tau} \Delta(\tau, E),
\end{equation}
where $\beta=1/T$, as usual. The mixed propagator $\Delta(\tau,
E)$, $0\le\tau\le\beta$, is given by
\begin{equation}\label{mixed-propagator}
\Delta(\tau, E)=\frac{1}{2E}\left[\left(1+n(E)\right)e^{-E\tau}+n(E)e^{E\tau}\right],
\end{equation}
where $n(E)=\left(e^{\beta E}-1\right)^{-1}$. For our purposes, it
will be convenient to use the following representations for the
mixed propagator \eqref{mixed-propagator}:
\ba
\Delta(\tau, E)&=&\frac{1}{2E}\left[1+n(E)(1+\mathcal{S}_{E})\right]
e^{-E\tau},
\label{mixed-propagator-operator-rep-a}
\\
&=&\frac{n(E)e^{\beta E}}{2E}
\left[1+e^{-\beta E}\mathcal{S}_{E}
\right]e^{-E\tau},
\label{mixed-propagator-operator-rep-b}
\ea
where $\mathcal{S}_{E}$ is the reflection operator defined in
\eqref{eq:def-reflection-operator}.
Substituting the representation
\eqref{mixed-propagator-operator-rep-a} back into
\eqref{propagator-saclay-rep} and using the fact that the operator
$(1/2E)[1+n(E)(1+\mathcal{S}_{E})]$ is linear, we obtain the
following equivalent Saclay representation for the scalar
propagator:
\ba\label{propagator-saclay-rep-2}
\Delta(k,E)=\frac{1}{2E}\left[1+n(E)(1+\mathcal{S}_{E})\right]
\int_0^\beta d\tau \,e^{(ik-E)\tau}.
\ea
\medskip

Consider now the two-vertex diagram with $\N$ internal lines of
figure \ref{fig:2-vertex}. Let $P=(p, \bp)$ its external
(incoming) 4-momentum (note that here $p$ stands for a
\emph{single} Euclidean energy variable) and let $K_j=(k_j=2\pi
Tn_j, \bk_j), j=1,\ldots,\N$, be the 4-momenta of the internal
lines, flowing from the left to the right vertex. The Matsubara
$D$-function corresponding to this graph is given by
\begin{equation}\label{two-vertex-D-function}
\begin{split}
D(p,\E, T)=&\gammaE\, T^{\N-1}\sum_{n_1, n_2, \dots, n_\N}\prod_{j=1}^\N \Delta(k_j, E_j)
\delta_{k_1+\cdots+k_\N\!,p},
\end{split}
\end{equation}
where the delta function is a Kronecker delta enforcing
conservation of energy at both vertices, $\sum_{j=1}^\N k_j=p$,
and $\gammaE$ was defined in \eqref{eq:def-gamma}.

Now, because the variables $p$ and $k_j$ are quantized in units of
$2\pi T$, the Kronecker delta in \eqref{two-vertex-D-function} can
be represented as
\ba
\delta_{k_1+\cdots+k_\N\!,p}=T\int_0^\beta d\tau
e^{-i\tau\left(k_1+\cdots+k_\N\!-p\right)},
\ea
so that the sums over the integers $n_j$ decouple:
\ba
D(p,\E, T)=\gammaE\, T^{\N}\int_0^\beta d\tau e^{ip\tau}
\prod_{j=1}^\N \left[\sum_{n_j}\Delta(k_j, E_j)e^{-i\tau
k_j}\right].
\ea
Using now the Saclay representation \eqref{propagator-saclay-rep-2}
for each propagator $\Delta(k_j,E_j)$ (with integration variable
$\tau_j$) we find
\begin{equation}
\begin{split}
D(p,\E, T)=&\;T^{\N} \prod_{j=1}^\N
\left[1+n(E_j)(1+\mathcal{S}_{E_j})\right]\\
&\times
\int_0^\beta d\tau e^{ip\tau}
\prod_{j=1}^\N
\left[
\int_0^\beta d\tau_j \,e^{-E\tau_j}
\sum_{n_j}e^{i(\tau_j-\tau) k_j}
\right].
\end{split}
\end{equation}
But
\ba\label{sac3.1}
T\sum_{n_j}e^{i(\tau_j-\tau) k_j}&=&\sum_{n}\delta(\tau_j-\tau +
n\beta)\nn\\
&=&\delta(\tau_j-\tau), \quad\text{for }0<\tau_j, \tau<\beta,
\ea
so that the final result for the Matsubara $D$-function for the
graph of figure \ref{fig:2-vertex} is:
\ba\label{two-vertex-D-function-result}
D(p,\E,T)&=&\prod_{j=1}^{I}
\left[1+n(E_j)(1+\mathcal{S}_{j})\right]
\int_0^\beta d\tau e^{(ip-E_{\rm tot})\tau}\nn\\
&=&\prod_{j=1}^{I}
\left[1+n(E_j)(1+\mathcal{S}_{j})\right]
\frac{e^{-\beta E_{\rm tot}}-1}{ip-E_{\rm tot}},
\ea
where $\mathcal{S}_{j}:=\mathcal{S}_{E_j}$, $E_{\rm
tot}:=\sum_{j=1}^{\N}E_j$ and we have used the fact that
$\exp(i\beta p)\equiv 1$.

\subsection{Proof of the Thermal Operator Representation}
We will now show that the result
\eqref{two-vertex-D-function-result} can be put into the form
\eqref{eq:conjecture-1}, where the zero-temperature $D$-function
for the graph of figure \ref{fig:2-vertex} is given by
\begin{equation}\label{two-vertex-D-zero-function-result}
D_0(p, \E)=-\left(\frac{1}{ip-E_{\rm tot}}
-\frac{1}{ip+E_{\rm tot}}\right),
\end{equation}
as can be easily be obtained from a calculation in the
old-fashioned perturbation theory formalism.
First, we observe that this function satisfies statement 3. In
fact, since the only cut set of the two-vertex diagram is the set of
all lines, we only need to show that the function
\eqref{two-vertex-D-zero-function-result} is annihilated by the
operator
\ba
\mathcal{A}:=\prod_{j=1}^\N(1+\mathcal{S}_{j}).
\ea
But since $\mathcal{S}_{x}$ is a reflection operator
($\mathcal{S}_{x}^2\equiv 1$) we have
\ba\label{DD1}
\left[\prod_{j=1}^{I}(1+\mathcal{S}_j)\right]\frac{1}{ip-E_{\rm tot}}
&=&\left[\prod_{j=1}^{I}(1+\mathcal{S}_j)\right]\left(\prod_{j=1}^{I}\mathcal{S}_j
\frac{1}{ip+E_{\rm tot}}\right),\nn\\
&=&\left[\prod_{j=1}^{I}(1+\mathcal{S}_j)\right]\frac{1}{ip+E_{\rm tot}},
\ea
so that indeed $\mathcal{A}D_0(p, \E)\equiv 0$. Therefore it is
enough to show that \eqref{eq:conjecture-1} holds with the thermal
operator in the form \eqref{eq:thermal-operator-form2}. But this
follows immediately from the following identity:
\ba\label{DD2}
\left[
\prod_{j=1}^{I}
\left[1+n(E_j)(1+\mathcal{S}_{j})\right]
\right]
\frac{e^{-\beta E_{\rm tot}}}{ip-E_{\rm tot}}
&=&
\left[
\prod_{j=1}^{I} n(E_j)e^{\beta E_j}\left[1+e^{-\beta E_j}\mathcal{S}_j\right]
\right]
\frac{e^{-\beta E_{\rm tot}}}{ip-E_{\rm tot}}\nn\\
&=&
\left[
\prod_{j=1}^{I} n(E_j)e^{\beta E_j}\left[1+e^{-\beta E_j}\mathcal{S}_j\right]
e^{-\beta E_j}\mathcal{S}_j
\right]
\frac{1}{ip+E_{\rm tot}}\nn\\
&=&
\left[
\prod_{j=1}^{I}\left[1+n(E_j)(1+\mathcal{S}_j)\right]
\right]
\frac{1}{ip+E_{\rm tot}},\nn
\ea
where we have used the property $\left(e^{-\beta
E_j}\mathcal{S}_j\right)^2\equiv 1$.
\section{The one-loop diagram}
\label{sec:one-loop-diagram}
\subsection{Calculation}\label{contour-method}
\begin{figure}
\centerline{\epsfig{file=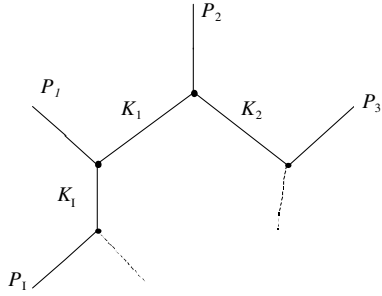,height=4cm,angle=0}}
\caption[]{A one-loop diagram with $I$ vertices.}
\label{fig:one-loop}
\end{figure}

The calculation of the Matsubara sum for the one-loop diagram with
$\N$ vertices and $\N$ internal propagators shown in figure
\ref{fig:one-loop} is most conveniently done using the standard
contour integration method\cite{books-on-FTQFT}. If a meromorphic
function $f$ has no singularities along the imaginary axis and $k$
stands for the Matsubara frequency $k=2\pi T n$, then
\begin{equation}\label{cont1}
T\sum_{n=-\infty}^{\infty} f(ik)=\frac{1}{2\pi i}\oint_C f(z)n(z) dz,
\end{equation}
where $n(z)=\left(e^{\beta z}-1\right)^{-1}$ and $C$ is the
positive contour that runs vertically upwards the complex
$z$-plane, infinitesimally to the right of the imaginary axis,
from $\varepsilon -i\infty$ to $\varepsilon + i\infty$, and then
comes back vertically and infinitesimally to the left of the
imaginary axis, from $-\varepsilon+i\infty$ to $-\varepsilon -
i\infty$, with $\varepsilon=0^+$.

If $|n(z)f(z)|$ goes fast enough to zero when $|z|$ goes to infinity
we can change the contour of integration to two negatively oriented
semicircumferences, one on each side of the imaginary axis, with
radii tending to infinity. Thus, by Cauchy's integral theorem,
\begin{equation}\label{cont2}
T\sum_{n=-\infty}^{\infty} f(ik)=-\sum_l {\rm Res}_{z=z_l}[f(z)n(z)],
\end{equation}
where $z_l$ are the poles of the function $f(z)$.
\medskip

Consider now the one loop graph of figure \ref{fig:one-loop}. Let
$P_i=(p_i, \mathbf{p}_i)$ be the external incoming momenta at each
vertex. Letting $k=2\pi nT$ be the Matsubara frequency of line
$1$, the Matsubara $D$-function in this case can be reduced to the
form
\begin{equation}\label{one-loop-D-function}
D(p, E, T)=\gammaE\,T\sum_n \Delta(k, E_1)\Delta(k+p_2, E_2)
\cdots\Delta(k+\sum_{j=2}^I p_j, E_I),
\end{equation}
where the energies $E_i$ are defined as before. Introducing a new
set of variables $u_j:=\sum_{l=1}^j p_l - p_1\;\;(j=2,\ldots,\N)$
and letting $u_1:=0$, we can write \eqref{one-loop-D-function} as
\begin{equation}\label{loop2}
D(p, E, T)=\gammaE\,T\sum_n \prod_{j=1}^I \Delta(k+u_j, E_j).
\end{equation}
Next using the identity
\begin{equation}\label{loop3}
\frac{1}{k^2+E^2}=-\frac{1}{2E}\left(\frac{-1}{ik+E}+\frac{1}{ik-E}\right)
=-\frac{1}{2E}\sum_{\sigma=\pm 1} \frac{\sigma}{ik-\sigma E},
\end{equation}
we get
\begin{equation}\label{loop4}
\begin{split}
D(p, E, T)=&(-1)^I \, T\sum_n \left\{ \prod_{j=1}^I
\left[\sum_{\sigma_j=\pm 1} \frac{\sigma_j}{ik+iu_j-\sigma_j E_j}\right]\right\},\\
=&(-1)^I \, T\sum_n \left\{\sum_{\sigma}
\prod_{j=1}^I\frac{\sigma_j}{ik+iu_j-\sigma_j E_j}\right\},
\end{split}
\end{equation}
where now $\sigma:=\{\sigma_1, \sigma_2, \dots, \sigma_I\}$.
If the function between brackets in \eqref{loop4} is called
$f(ik)$, then we see that the poles of $f(z)$ are located at
$z_l=-iu_l+\sigma_l E_l$, so that the application of \eqref{cont2}
gives us
\begin{equation}\label{loop5}
D(p, E, T)=(-1)^{I+1} \,
\sum_{\sigma} \sum_{l=1}^{I}n(\sigma_l E_l)\sigma_l \prod_{j\not{=} l}^I
\frac{\sigma_j}{i(u_j-u_l)+\sigma_l E_l -\sigma_j E_j}.
\end{equation}
In order to express the result for the $D$-function in terms of
Bose-Einstein factors of positive argument only, we will perform
the summation over $\sigma_l$ explicitly. Introducing the notation
$\sigma_{\lnot}:=(\sigma_1, \dots, \sigma_{l-1},
\sigma_{l+1}, \dots, \sigma_I)$ and using the identity
$n(-E_l)=-(1+n(E_l))$ we find
\begin{equation}\label{loop6}
\begin{split}
D(p, &E, T)=(-1)^{I+1}
\sum_{l=1}^{I} \sum_{\sigma_{\lnot}}\left\{\prod_{j\neq l}^I
\frac{\sigma_j}{i(u_j-u_l)-E_l -\sigma_j E_j}+\right.\\
&\phantom{xxxxxxxxxxxxxxxxxxxx}
\left.+n(E_l)\left[\prod_{j\neq l}^I
\frac{\sigma_j}{i(u_j-u_l)-E_l -\sigma_j E_j}+\prod_{j\neq l}^I
\frac{\sigma_j}{i(u_j-u_l)+E_l -\sigma_j E_j}\right]\right\}.
\end{split}
\end{equation}
In terms of the auxiliary function
\begin{equation}\label{loop7}
\begin{split}
d_{l}(p, E):=&\sum_{\sigma_{\lnot}}\prod_{j\neq l}^I
\frac{\sigma_j}{i(u_j-u_l)-E_l -\sigma_j E_j}\\
=&\prod_{j\neq l}^I\sum_{\sigma_j}
\frac{\sigma_j}{i(u_j-u_l)-E_l -\sigma_j E_j},
\end{split}
\end{equation}
and the reflection operator, $\mathcal{S}_{i}:=\mathcal{S}_{E_i}$,
defined in \eqref{eq:def-reflection-operator} we have
\begin{equation}\label{one-loop-D-function-result}
D(p, E, T)=(-1)^{I+1}\sum_{l=1}^{I} \left[
d_{l}(p, E)+n(E_l)(1+\mathcal{S}_{E_l})d_{l}(p, E)\right].
\end{equation}
\subsection{Proof of the Thermal Operator Representation}

We shall prove now that the $D$-function
\eqref{one-loop-D-function-result} for the one-loop graph of
figure \ref{fig:one-loop} can be written in the form
\eqref{eq:conjecture-1}, as
\begin{equation}\label{one-loop-conjecture-holds}
D(p, E, T)=\left[1+\sum_{j=1}^{I} n(E_j)(1+
\mathcal{S}_{j})\right]D_0(p, E),
\end{equation}
with
\begin{equation}\label{one-loop-D-zero-function}
D_0(p, E)=(-1)^{I+1}\sum_{l=1}^{I}
d_{l}(p, E).
\end{equation}
Since the graph of figure \ref{fig:one-loop} gets disconnected if
two or more lines are snipped, the thermal operator has terms no
higher than linear in the Bose-Einstein factors $n(E)$. But
equation \eqref{one-loop-conjecture-holds} will reduce to eqn.
\eqref{one-loop-D-function-result} if the operator
$(1+\mathcal{S}_{j})$ annihilates the auxiliary function
$d_l(p, E)$ when $j\neq l$. This is indeed the case:
from equation \eqref{loop7} we see that when $j\neq l$
\begin{equation}\label{loop10}
\begin{split}
d_{l}(p, E)=&\sum_{\sigma_j}\frac{\sigma_j}{i(u_j-u_l)-E_l -\sigma_j E_j}
\prod_{k\neq l\!,\,j}^I\sum_{\sigma_k}
\frac{\sigma_k}{i(u_k-u_l)-E_l -\sigma_k E_k},\\
=&\sum_{\sigma_j}\frac{-\sigma_j}{i(u_j-u_l)-E_l +\sigma_j E_j}
\prod_{k\neq l\!,\,j}^I\sum_{\sigma_k}
\frac{\sigma_k}{i(u_k-u_l)-E_l -\sigma_k E_k}\\
=&-\mathcal{S}_j d_{l}(p, E),
\end{split}
\end{equation}
which means that
\begin{equation}\label{loop11}
(1+\mathcal{S}_j)d_{l}(p, E)\equiv 0,\quad\text{if
}j\neq l.
\end{equation}
Statement 1 is then valid for the one-loop graph of figure
\ref{fig:one-loop}.

Furthermore, statement 3 is also true for this graph.
In fact, any cut set will at least contain two lines, say
lines $i$ and $j$. But then
\begin{equation}
(1+\mathcal{S}_{i})(1+\mathcal{S}_{j})d_{l}(p, E)
\equiv 0,
\end{equation}
since, for any given $l$, either $i$ or $j$ will be different
from $l$ (since $i\neq j$), leading to a vanishing
contribution because of \eqref{loop11}.

\section{Further evidence and conclusions}
\label{sec:conclusions}

One piece of evidence in favor of the general validity of the
representation \eqref{eq:conjecture-1} is provided by a comparison
with a well known result of thermal field theory, first formulated
by Weldon\cite{weldon}, concerning the interpretation of the
imaginary part of the retarded self-energy $\Pi_R$ in terms of the
direct and inverse decay rates of a particle propagating in the
thermal medium. A well-known result of quantum statistical
mechanics\cite{Baym and Mermin} is that the full retarded
self-energy $\Pi_R$ can be obtained from the Euclidean self-energy
$\Pi_\beta$ by analytic continuation as
\ba
\Pi_R(\omega,{\mathbf{p}})
=-\Pi_\beta(i(\omega+i\varepsilon),{\mathbf{p}}),
\ea
where $\omega$ stands for a real continuous variable. In the
context of perturbative quantum field theory, the imaginary part
of $\Pi_R$ is given in the form of integrals over phase space of
amplitudes squared, weighted by certain statistical factors that
account for the possibility of particle absorption from the medium
or particle emission into the medium\cite{weldon}. For example,
for the one-loop 2-vertex diagram corresponding to figure
\ref{fig:OFPT-diagrams} in the appendix that follows, the result
for the imaginary part of the retarded self- energy is (we have
set $g\equiv 1$)
\ba\label{eq:example-imaginary-part}
  \operatorname{Im} \Pi _R (\omega ,{\mathbf{p}}) &=&  - \pi \int {\frac{{d^3 k}}
{{(2\pi )^3 }}\frac{1}
{{4E_1 E_2 }}}
  \Big[ {(1 + n_1  + n_2 )}\big( {\delta (\omega  - E_1  - E_2 ) - \delta (\omega  + E_1  + E_2 )} \big) \nn\\
  &&\phantom{xxxxxxxxxxxxxxxx}
  { - (n_1  - n_2 )\big( {\delta (\omega  - E_1  + E_2 )
  - \delta (\omega  + E_1  - E_2 )} \big)} \Big].
\ea
where $n_i\equiv n(E_i)$.
But from the general form \eqref{eq:D-def} for a diagram in the
Euclidean formalism, it is clear that the imaginary part of the
analytically continued diagram is determined by the analytic
continuation of its $D$-function. The general validity of our main
representation in the form \eqref{eq:conjecture-1} would imply
that the latter is in turn completely determined in terms of the
analytic continuation of the zero-temperature $D$-function, $D_0$,
since the thermal operator $\hat O$ is real and does not involve
the external momenta.

For the particular simple diagram we are considering, which is
actually a special case of the general 2-vertex graph considered
in section \ref{sec:two-vertex-diagram}, the Thermal Operator
Representation has been proven to hold. Hence
\ba\label{eq:consequence-of-conjecture}
\operatorname{Im} D\left( {i(\omega  + i\varepsilon ),E_1 ,E_2 ,T}
\right) = \hat O\left( {E_1 ,E_2 ,T} \right)\operatorname{Im} D_0
\left( {i(\omega  + i\varepsilon ),E_1 ,E_2 } \right).
\ea
The last imaginary part could in principle be obtained from the
standard cutting rules that apply in zero-temperature field
theory, without having to compute $D_0$ itself. In this case,
however, we have the closed result
\eqref{two-vertex-D-zero-function-result} for $D_0$, which allows
us to compute directly
\ba
\operatorname{Im} D_0
\left( {i(\omega  + i\varepsilon ),E_1 ,E_2 } \right)&=&
\operatorname{Im} \left[ {\frac{1}
{{\omega  + E_1  + E_2  + i\varepsilon }} - \frac{1}
{{\omega  - E_1  - E_2  + i\varepsilon }}} \right]\nn\\
&=&
- \pi \left[ {\delta (\omega  + E_1  + E_2 ) - \delta (\omega  - E_1  - E_2 )}
\right].
\ea
Now in this case the thermal operator is given by
\ba\label{eq:thermal-operator-example}
\hat O\left( {E_1 ,E_2 ,T} \right) = 1 + n_1 (1 + \mathcal{S}_1 ) + n_2 (1 + \mathcal{S}_2 )
= 1 + n_1  + n_2  + n_1 \mathcal{S}_1  + n_2 \mathcal{S}_2.
\ea
Since
\[
n_1 \mathcal{S}_1 \left[ {\delta (\omega  + E_1  + E_2 ) - \delta (\omega  - E_1  - E_2 )} \right]
= n_1 \left[ {\delta (\omega  - E_1  + E_2 ) - \delta (\omega  + E_1  - E_2 )}
\right],
\]
etc., we readily obtain
\ba
\hat O \operatorname{Im} D_0 &=&
\pi\Big[{(1 + n_1  + n_2 )}\big( {\delta (\omega  - E_1  - E_2 ) - \delta (\omega  + E_1  + E_2 )} \big) \nn\\
  &&\quad { - (n_1  - n_2 )\big( {\delta (\omega  - E_1  + E_2 )
  - \delta (\omega  + E_1  - E_2 )} \big)} \Big],
\ea
thereby reproducing \eqref{eq:example-imaginary-part}, with all
the correct signs and thermal factors.
\bigskip

In this paper we have restricted our attention to some simple
diagrams in the finite-temperature imaginary-time formalism for a
scalar relativistic field theory. We have shown that the full
result of performing the Matsubara sum associated to any given
Feynman graph can be obtained from its zero-temperature
counterpart, by means of a simple linear operator. Given the
general form \eqref{eq:thermal-operator-form1} of the thermal
operator, which can be readily and naturally extrapolated to
diagrams of arbitrary topologies, it is not at all implausible
that the representation \eqref{eq:conjecture-1} be actually valid
in complete generality. This generalization remains an open
problem however, and work in this direction is in progress.

An analysis similar to the one presented here should apply in a
theory containing fermions; the algebra will be slightly more
complicated because of the spin structure. We have deferred this
analysis, as well as the extension of our results to gauge
theories, until we have been able to prove or disprove that the
Thermal Operator Representation put forward in this paper does
indeed hold for an arbitrary loop graph in a scalar field theory.

\section{Acknowledgement}
The authors would like to thank C.~Dib and I.~Schmidt for
suggestions, O.~Orellana for interesting insights, and the first
referee for helpful criticisms and comments to the original
version of this work.
This work was supported by CONICYT, under grant Fondecyt 8000017.
E.S. would also like to thank the Physics Department of
Universidad Santa Mar{\'\i}a for its hospitality and the partial
financial support of the Millennium Scientific Nucleus ICM
P99-135F.

\section{Appendix}
\subsection{The OFPT-rules}
\label{ofpt-rules-statement}
\begin{figure}[b]
\centerline{\epsfig{file=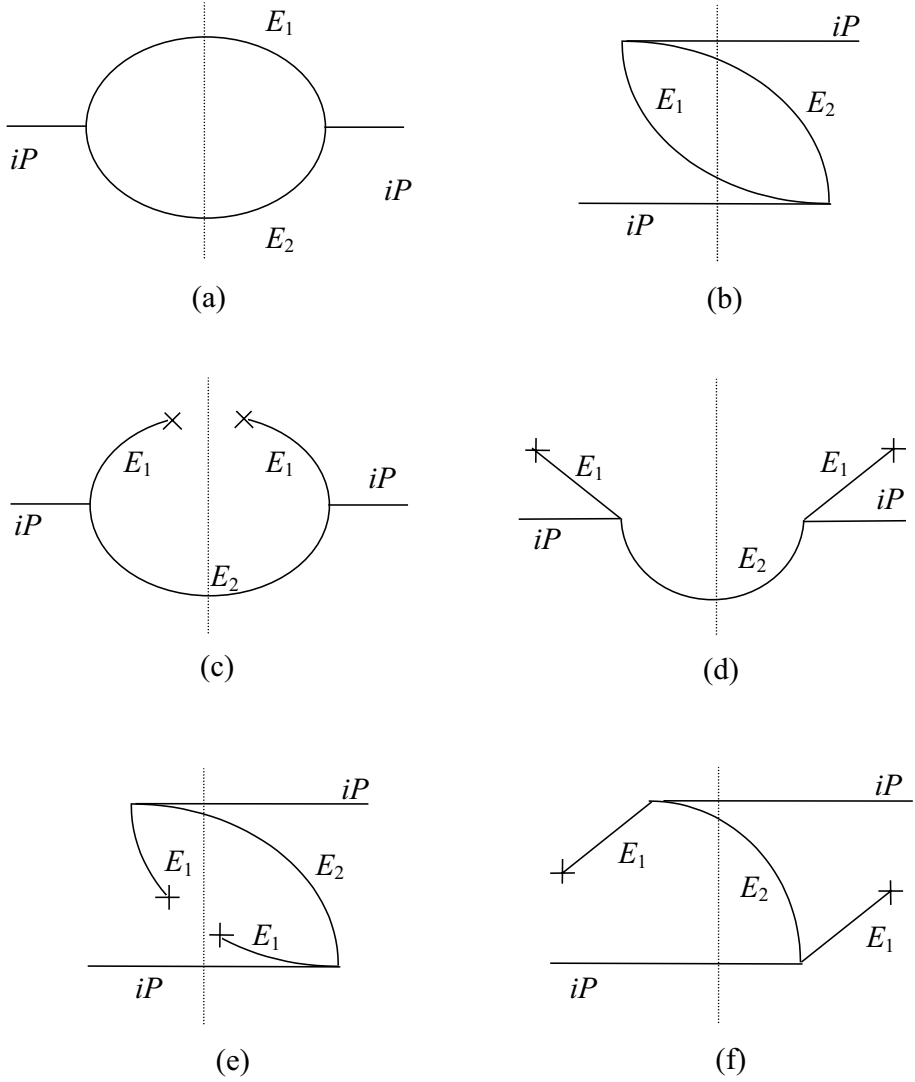,height=16cm,angle=0}}
\caption[]{An example of the diagrams which appear in the OPFT-rules.}
\label{fig:OFPT-diagrams}
\end{figure}

The rules originally put forward in reference \cite{DES-97} to
write down an explicit expression for the Matsubara $D$-function
corresponding to the general scalar graph considered in section
\ref{sec:conjecture} are given by the following statements
(it might be useful to refer to figure~\ref{fig:OFPT-diagrams} at
this point):
\begin{list}
{\alph{bean}.}{\usecounter{bean}}
\item For each external line, characterized by a real Euclidean
4-vector $(p_{l}, {\bf p}_l)$, define its energy as $ip_{l}$. For
each internal line define its energy as $E_i=({\bf
k}_i^2+m_i^2)^{1/2}$, where ${\bf k}_i$ is the 3-momentum carried
by the line and $m_i$ is the mass of the propagating particle.
\item
Define a {\em direction of time} or {\em energy flow} (which we
shall take conventionally from left to right) and consider all
possible orderings of the vertices along this direction (see,
e.g., Figs.~\ref{fig:OFPT-diagrams}.a and
\ref{fig:OFPT-diagrams}.b. For a graph with $n+1$ vertices there
will be $(n+1)!$ such orderings).
\item
For each time-ordered graph generated in (b) consider, in addition
to itself, all possible {\em connected} graphs that can be
obtained by snipping any number of internal lines. Each line that
is snipped becomes a pair of legs we shall call {\em thermal
legs}. Attach a cross to their ends to distinguish them from the
original external lines of the graph. Both legs of a given pair
inherit the energy $E_i$ of the internal line that originated
them. However, one leg must be oriented as {\em incoming} with
energy $E_i$ and the other as {\em outgoing} with energy $E_i$.
Both possible orientations have to be considered, each one
generating a different diagram (see, e.g., Figs.
\ref{fig:OFPT-diagrams}.c and \ref{fig:OFPT-diagrams}.d).
\item
For each graph in (c), define its total incoming energy,
$E_{inc}$, as the sum of all incoming external energies plus the
energies of all incoming thermal legs that join the diagram {\em
before} their outgoing partner (e.g. as in Figs.
\ref{fig:OFPT-diagrams}.d and \ref{fig:OFPT-diagrams}.f). Thermal
leg pairs that satisfy this property shall be referred to as {\em
external}, and those that do not as {\em internal} (e.g. as in Figs.
\ref{fig:OFPT-diagrams}.c and \ref{fig:OFPT-diagrams}.e). Then,
associate to this graph an expression equal to the product of the
following factors:
\begin{list}
{\arabic{pea}.}{\usecounter{pea}}
\item
draw a full vertical division (a ``cut'') between each pair of
consecutive time-ordered vertices (there are $n$ such cuts in a
graph with $n+1$ vertices); for each cut include a factor
\begin{equation}
\frac {1}{E_{inc}-E_{cut}}
\label{eq:energy-denominator}
\end{equation}
where $E_{cut}$ is the total energy of the intermediate state
associated with the cut, defined as the sum of the energies of all
the lines that cross the cut in question (as in zero-temperature
time-ordered perturbation theory), plus the energies of all {\em
internal} thermal pairs whose originating internal line would have
crossed the cut.
\item
include a thermal occupation factor $n_i\equiv n(E_i)$ for each
thermal pair (of energy $E_i$) in the diagram (if any).
\item
include an overall factor of $(-1)^n$, where $n+1$ is the number
of vertices.
\end{list}
\item
The integrand $D(p, E, T)$ in \eqref{eq:D-def}, \ie the
Matsubara $D$-function of the graph, is the
sum of the expressions computed according to rule (d), over all
the graphs in (c).
\end{list}
\subsection{An algebraic approach to the OFPT-rules}
\label{operator-approach}
Let us call $D_R(p, E, T)$ the expression for the
$D$-function generated according to the OFPT-rules. A trivial
check the OFPT-rules do satisfy is that they yield the known
correct result in the limit $T\to 0$, keeping the external Euclidean
energies $p$ fixed. In fact, in the limit $T\to 0$ all the
thermal factors $n(E)$ vanish, so that according to the rules
$D_R(p, E, 0)$ is just given by all possible time
ordered diagrams with no snipped lines, calculated according to
rule (d) above. But this is precisely the result one would obtain
calculating the $T=0$ Euclidean graph (with external momenta
$(p_l, \bp_l)$) using old-fashioned perturbation
theory\cite{OFPT}. We have therefore:
\begin{equation}\label{eq:proof-zero-T}
D_0(p, E)=D_R(p, E, 0),
\end{equation}
where $D_0(p, E)$ is the $D$-function associated
with the zero-temperature Euclidean Feynman graph.
Hence, the rules hold at $T=0$.

At finite temperature, we get extra contributions according to
rules (c) and (d) above. Now, instead of considering, as commanded
by rule (c), all possible connected graphs that can be obtained by
snipping \emph{any} number of internal lines of a \emph{given}
``un-snipped'' time-ordered graph, let us rather group the snipped
diagrams according to \emph{which} lines are snipped, regardless
of the time-ordering. Take for instance all the diagrams which
have only the $i$-th line snipped ($i$ is fixed). A set of this
type is conformed, for instance, by the diagrams
(c) to (f) of Fig.\ref{fig:OFPT-diagrams}. It follows
directly from rule (d.1) that a diagram in which the snipped line
forms an \emph{internal} thermal leg pair (\ie we have a
``closed'' snipping, as in Figs.~\ref{fig:OFPT-diagrams}.c and
\ref{fig:OFPT-diagrams}.e) has exactly the same mathematical
weight as the zero-temperature ``un-snipped'' diagram, except of
course for the extra thermal factor $n(E_i)$. Thus the sum of all
these diagrams, \ie the diagrams that have only the $i$-th line
snipped closed, adds up to $n(E_i)D_0(p, E)$. On the
other hand, if the snipped line forms an \emph{external} thermal
leg pair (\ie we have a ``open'' snipping, as in
Figs.~\ref{fig:OFPT-diagrams}.d and \ref{fig:OFPT-diagrams}.f), we
again have an extra thermal factor $n(E_i)$, but now the rest of
the expression differs from that for the ``un-snipped'' graph in
the sign of the energy $E_i$. This is so because, for an open
snipping, the energy $E_i$ moves from $E_{cut}$ to $E_{inc}$, as
can be gathered from rule (d).

Let $x$ symbolize a variable, and let $\mathcal{S}_x$ be the
operator that acts on functions of $x$, changing the sign of the
argument $x$, according to:
\[
\mathcal{S}_x f(x) := f(-x).
\]
In terms of the reflection operator $\mathcal{S}_x$, we can write
the sum of all the time-ordered diagrams with only the $i$-th line
snipped open as
\[
n(E_i)\mathcal{S}_i D_0(p, E),
\]
where we have written $\mathcal{S}_i:=\mathcal{S}_{E_i}$ to avoid
cluttering the notation. So the full contributions of the diagrams
in which only the $i$-th line is snipped can be written as
\[
n(E_i)\left(1+\mathcal{S}_i\right) D_0(p, E).
\]
\medskip

The analysis above can clearly be generalized to add up the
contribution of the graphs with more than one snipped line. Taking
into account that only connected graphs are allowed by the
OFPT-rules (so that one is allowed to snip at most $L$ internal
lines, where $L$ is the number of independent loops), we arrive at
the following result:
\begin{theorem}\label{thm:rules-math-form}
The OFPT-rules admit the following mathematical representation:
\begin{equation}\label{eq:rules-math-form1}
D_R(p, E, T)=\hat{O}(E, T)D_0(p, E),
\end{equation}
where $\hat{O}(E, T)$, the thermal operator, is given by
\begin{equation}\label{eq-appendix:thermal-operator-form1}
\begin{split}
\hat{O}(E,T):=1+&\sum_{i=1}^I  n(E_i)(1+\mathcal{S}_i)
+\sideset{}{^\prime}\sum_{\langle i_1,  i_2\rangle} n(E_{i_1})n(E_{i_2})
(1+\mathcal{S}_{i_1})(1+\mathcal{S}_{i_2})\\
+& \dots + \sideset{}{^\prime}\sum_{\langle i_1, \dots,  i_{L}\rangle}
\prod_{l=1}^{L} n(E_{i_l})(1+\mathcal{S}_{i_l}).
\end{split}
\end{equation}
Here the indices $i_1, i_2, \ldots$ run from 1 to $\N$ (the number
of internal propagators) and the symbol $\langle i_1, \dots,
i_k\rangle$ stands for an unordered $k$-tuple with no repeated
indices. The primes on the summation symbols imply that we are to
exclude from the sums those tuples $\langle i_1, \dots,
i_k\rangle$ such that if we snip all the corresponding lines
$i_1,\dots,i_k$ then the graph becomes disconnected.
\end{theorem}

\newcommand{\xprd}[3]{Phys.~Rev.~{\bf D#1}, #3 (#2)}
\newcommand{\xprl}[3]{Phys.~Rev.~Lett.~{\bf #1}, #3 (#2)}
\newcommand{\xpr}[3]{Phys.~Rev.~{\bf #1}, #3 (#2)}
\newcommand{\plb}[3]{Phys.~Lett.~{\bf B#1}, #3 (#2)}
\newcommand{\pla}[3]{Phys.~Lett.~{\bf A#1}, #3 (#2)}
\newcommand{\hepth}[1]{hep-th/#1}
\newcommand{\hepph}[1]{hep-ph/#1}
\newcommand{\condmat}[1]{cond-mat/#1}
\newcommand{\mpla}[3]{Mod.~Phys.~Lett.~{\bf A#1}, #3 (#2)}
\newcommand{\jhep}[3]{JHEP~{\bf #1}, #3 (#2)}
\newcommand{\xrmp}[3]{Rev.~Mod.~Phys.~{\bf #1}, #3 (#2)}
\newcommand{\jmp}[3]{Jour.~Math.~Phys.~{\bf #1}, #3 (#2)}
\newcommand{\npb}[3]{Nucl.~Phys.~{\bf B#1}, #3 (#2)}
\newcommand{\epjc}[3]{Eur.~Phys.~J.~{\bf C#1}, #3 (#2)}

\end{document}